\documentclass[a4paper,fleqn,usenatbib]{mnras}

\usepackage[T1]{fontenc}
\usepackage{ae,aecompl}

\usepackage{graphicx}	% Including figure files
\usepackage{amsmath}	% Advanced maths commands
\usepackage{amssymb}	% Extra maths symbols
\usepackage{xcolor}
\title[LMXB]{Spectroscopic modelling of four neutron star low-mass X-ray binaries using CLOUDY}

\author[G. Shaw et al.]{
Gargi Shaw$^{1}$,\thanks{E-mail: gargishaw@gmail.com (GS)}
Sudip Bhattacharyya$^{1}$
\\
$^{1}$Department of Astronomy and Astrophysics, Tata Institute of Fundamental Research, \\ Homi Bhabha Road, Navy Nagar, Colaba, 
Mumbai 400005, India\\
}

\date{Accepted for publication in MNRAS}

\pubyear{2019}

\begin{document}
\label{firstpage}
\pagerange{\pageref{firstpage}--\pageref{lastpage}}
\maketitle

% Abstract of the paper
\begin{abstract}
Low-mass X-ray binaries (LMXBs) have a wide range of X-ray properties
which can be utilised to reveal many physical conditions of the associated accretion
discs. We use the spectral
synthesis code CLOUDY to perform a detailed modelling of neutron star LMXBs GX 13+1, MXB 1659--298, 4U 1323--62 and XB 1916--053; and 
characterise the underlying physical conditions, such as density, radiation
field, metallicity, wind velocity, etc. For this purpose we model highly ionised spectra of Fe, Ca, S, Si, Mg, Al in the soft X-ray band, and compare 
the predicted line flux ratios with the observed values. We 
also find that the strength and profile of these spectral lines get modified in the presence of magnetic field in the accretion disc. Using this, we estimate an upper limit of 
the existing magnetic field to be about a few hundred to a few thousand G in the accretion discs of these four LMXBs.
\end{abstract}

% Select between one and six entries from the list of approved keywords.
% Don't make up new ones.

%%%%%%%%%%%%%%%%%%%%%%%%%%%%%%%%%%%%%%%%%%%%%%%%%%

%%%%%%%%%%%%%%%%% BODY OF PAPER %%%%%%%%%%%%%%%%%%

\section{Introduction}
A low-mass X-ray binary (LMXB) system consists of a neutron star or a black hole, and a companion star with a mass
similar to or less than the Sun. The companion star transfers its mass by Roche-lobe overflow, in which the material initially gets 
pulled into an accretion disc around the neutron star or the black hole and then slowly spirals 
into the enormous gravitational well of the compact object. The accreted material is heated up to a very high temperature 
causing the system to shine brightly in X-rays \citep{Shapiro1984, Tauris2006}.

High-inclination LMXBs usually show absorption lines in the X-ray spectrum, as the line-of-sight passes through the structures above the accretion 
disc in such 
cases. In fact iron K$\alpha$ fluorescence lines, Fe XXV (6.700 keV) and Fe XXVI (6.966 keV), have been observed from many high-inclination LMXBs, such as 
X 1254--690, MXB 1658--298, X 1624--490
\citep{Trigo2006}, XTE J1710--281 \citep{2009A&A...502...905} and 4U 1916--053 \citep{Trigo2006}. In addition to these Iron K$\alpha$ lines, many other highly 
ionised lines for Mg XII (1.472 keV), Al XIII (1.728 keV), Si XIV (2.005 keV), S VI (2.621 keV), Ca XX (4.105keV), Ne X (1.0218 keV), O VIII (0.6536 keV) 
have been observed from some LMXBs, such as GX 13+1 \citep{D'Ai2014} and MXB 1659--298 \citep{2018Iariab}. 
These lines contain a wealth of information, and hence can be used to decipher the underlying physical conditions which are responsible 
for their origin. In this paper we aim to study four neutron star LMXBs, GX 13+1, MXB 1659--298, 4U 1323--62 and XB 1916--053, by detailed spectroscopic modelling.

The LMXB 4U 1323--62 was first observed with \citep{vanderKlis1985} the UHURU satellite. It has
the orbital period of 2.94 hour \citep{2005Boirin} (hereafter B05) and shows 1 Hz quasi-periodic oscillation (QPO) and frequent
thermonuclear bursts. Simultaneous intensity dips and bursts in X-rays were also observed from this source \citep{vanderKlis1985}. Later, B05 observed
highly ionised Fe
XXV and Fe XXVI lines at 6.68 $\pm$ 0.04 keV and 6.97$\pm$0.05 keV respectively, and
also the change in intensity for the same lines, with  XMM--Newton. They 
estimated the angle between the line-of-sight
and the rotational axis of the accretion disc to be
approximately 60 degrees.  On the other hand, XB 1916--053 was first
observed by \citet{1982ApJ...253...L61} with a orbital period of 3000.6 $\pm$0.2 sec \citep{2001MNRAS...322...827}. 
It is 9.3 kpc away
\citep{2010MNRAS...401...1275} and the angle between the line-of-sight
and the rotational axis of the accretion disc is approximately 77
degrees. MXB 1659--298 was first observed by \citet{1976Lewin}. Recently, \citet{2018Iariab} have observed several highly ionised lines, such as 
Ne X, O VIII, Fe XXV and Fe XXVI, and estimated an inclination angle of 72 $\pm3$ degrees \citep{2018Iariaa}. 
GX 13+1 is another important LMXB which shows many highly ionised absorption lines. It is one of the brightest Galactic LMXBs, which is at a distance of 7$\pm$1 kpc. It has the longest known orbital period for a Galactic
neutron star LMXB, and it has a K5III spectral type donor star \citep{Bandyopadhyay1999}. GX 13+1 was first observed by \cite{Fleischman1985} using HEAO 1 
satellite. Later a periodicity of 24.27 days was observed in its power spectrum density obtained from data collected over 14 years with the All Sky Monitor (ASM) on board the Rossi X-ray Timing Explorer (RXTE) \citep{Iaria2013}. Though it has been classified as an atoll source, it shows properties that are close to the Z sources and several properties still remain unexplained. Moreover, GX 13+1 shows thermonuclear bursts and super bursts in X-ray, and it also shows a QPO at 61$\pm 1$ Hz \citep{D'Ai2014}. \citet{D'Ai2014} have also observed dips for a duration of 450 secs. The inclination angle of the source is > 65$^{\circ}$. 
A wealth of X-ray spectroscopic observations is available for GX 13+1, which includes Fe XXV, Fe XXVI, Mg XII, Al XIII, 
Si XIV, S VI, Ca XX  absorption lines in the persistent phase \citep{D'Ai2014}. 
  
These absorption lines from the above mentioned sources are useful to decipher the chemical, physical as well as
kinematic properties of the accretion structure.
We use the spectral simulation code, CLOUDY, in this work to model iron K$\alpha$ fluorescence lines, Fe XXV  and Fe XXVI, from two LMXBs, 4U 1323--62 
and XB 1916--053 and also other highly ionised lines like Mg XII, Al XIII, Si XIV, S XVI , Ca XX, Ne X, O VIII from GX 13 +1 \citep{D'Ai2014} and 
MXB 1659--298 \citep{2018Iariab}. 
We mainly focus on spectroscopic modelling of these highly ionised lines in the soft X-ray band and determine the underlying physical conditions 
which give rise to these lines.

Moreover, such a large number of absorption lines could also be analysed in a way such that they can help to probe the magnetic field in the 
accretion disc. So far, to our best knowledge, no one has estimated the strength of magnetic field or its upper limit in the accretion discs  
of 4U 1323--62, XB 1916--053, MXB 1659--298 and GX 13+1. We take this opportunity to constrain the magnetic fields for these accretion discs.  

Here, we mention some important roles of disk magnetic field. In accretion discs, magnetic field is an important source of viscosity. Weakly-magnetised 
differentially-rotating accretion discs are generally 
unstable and shows magnetohydrodynamical instability and magnetohydrodynamical turbulence \citep{Balbus1994}, which could give rise to outbursts. 
Many accreting systems show powerful jets 
and it is believed that  the magnetic field couples discs and jets in such systems.
Highly collimated jets in black hole binaries could be explained by strong magnetic field existing in the inner accretion disc \citep{McKinney2007}. 
However, the origin of accretion disc  magnetic field is not well understood yet. In case of an accreting neutron star with a magnetosphere, it is very natural to expect that the
surrounding accretion disc will also harbour some amount of magnetic
field \citep{2013MNRAS...435...2633}. This may be due to threading of
neutron star magnetosphere and its own magnetic field. \citet{2016ApJ...822...33} numerically showed that the interaction of neutron star magnetic field 
with its differentially rotating accretion disc gives rise 
to opening of pulsar magnetic field lines. This could be tested if the disc magnetic field can somehow be measured. It is thus very important to estimate 
the magnetic field in accretion discs. 
\citet{Silantev2013} have estimated magnetic fields 
of active galactic nuclei (AGN) using polarised broad H$\alpha$ lines. They used wavelength dependence of polarisation degree 
inside this line to measure magnetic field, and found 
B$_\parallel$= 0, B$_\phi$=14 G for the AGN Akn 120 and B$_\parallel$=0.6 G, B$_\phi$ = 8.5 G for Mrk 6, respectively. Linear polarimetric observations in 
near-infrared and optical band of LMXB jets were done by \citet{Russel2018}, but they 
did not calculate magnetic field strength.
In fact, so far, to the best of our knowledge, the accretion disc magnetic field for LMXBs has not been measured directly.
The strength of this field is not expected to be high enough to produce a detectable Zeeman splitting. In future, X-ray polarimetry 
could be used to estimate such a field.

In this work, along with modelling the observed lines in the soft X-ray band to determine physical parameters of the accretion discs, 
we also show that it is possible to estimate an upper limit of the magnetic field in an accretion disc using the absorption lines. 
We apply this method for 4U 1323--62, XB 1916--053, MXB 1659--298 and GX 13+1.

This paper is organised as follows. In
section 2, we describe our calculations. Results for the LMXBs  4U 1323--62, 4U 1916--053, MXB 1659--298, GX 13+1, and our conclusions are presented 
in sections 3 and 4 respectively.
\section{Calculations}

\subsection{Numerical Methods}
\label{sec:Calculations} 
In this section, we present our calculations with the detailed method which are carried out using the spectral synthesis code 
CLOUDY \citep{Ferland2013, Ferland2017}. It is a micro-physics code based on a self-consistent $\textit {ab initio}$ calculation of thermal, 
ionisation, and chemical balance of non-equilibrium 
gas and dust 
exposed to a source of radiation. This software and its documentation is freely available at {\it http://nublado.org}, and it is widely used as 
one of the astrophysical plasma codes.
It can be utilised to model emission lines in various astrophysical environments, 
from gaseous nebulae to quasars covering a wide range of temperature and 
density. It predicts column densities of various species and resultant spectra  
ranging from  gamma rays to radio and vice versa using a minimum number of input parameters like density, radiation field, chemical composition and geometry.
 There are various options 
to make a simulated model to the most realistic one. Currently CLOUDY includes 625 species
including atoms, ions and molecules and uses five
distinct databases: H-like and
He-like iso-electronic sequences \citep{2012Porter}, the H$_{2}$ molecule \citep{2005Shaw}, Stout, CHIANTI \citep{2012Landi} and
LAMBDA \citep{2005Schoier} to model spectral lines. Atoms of the H-like iso-electronic sequence have one bound electron and
atoms of the He-like iso-electronic sequence have two
bound electrons. CLOUDY uses an unified
model for both the H-like and He-like iso-electronic sequences, that extends from H to Zn, as described by \citet{2012Porter}. The highly ionised lines modelled here belong to either H-like or He-like iso-electronic sequences. O VIII, Ne X, Ca XX, Fe XXVI, S XVI, Si XIV, Mg XII, Al XIII have H-like iso-electronic sequences, whereas Fe XXV has He-like iso-electronic sequence. Any number of levels up to 400 can be computed and the ionisation and level populations are self-consistently determined by solving the full collisional-radiative problem. Increasing the number of levels allows a better representation of the collision physics that occurs within higher levels of the atom but it consumes more time and memory. In the default mode of CLOUDY, levels up to 50 are included. We have included default number of levels in our calculations.
Earlier, various research groups, for example, B05 and \citet{2015MNRAS...453...292},  used CLOUDY to model LMXBs. Here we aim to use the version c-17 of CLOUDY which is more advanced than the 
earlier versions and our models incorporate more parameters than that were included previously which will be discussed in this section. 

For all of our models, we assume a geometry of a constant pressure clumpy highly-ionised gaseous rotating disc around a 1.4 solar mass 
neutron star with high velocity wind. Earlier groups have neither included clumpiness nor the rotating disc around a 1.4 solar mass 
neutron star with high velocity wind. Here hydrogen density, incident radiation field, metallicity, clumpiness and wind are model parameters 
whose values can be varied. Below we discuss our input parameters briefly. 

For hydrogen density $n_{H}$ (cm$^{-3}$), we use the total hydrogen density and is given by 
\begin {equation}
n_{H}=n_{H_{0}}+n_{H^{+}}+2n_{H_{2}}+\sum_{other}n_{H_{other}},      
\end {equation}
where $n_{H_{0}}$, $n_{H^{+}}$, $2n_{H_{2}}$ and $n_{H_{other}}$ represent H in neutral, ionised, hydrogen molecule and all other hydrogen-bearing molecular states, 
respectively. One can vary the density spatially in 
different ways across the whole gaseous extent. We use following radius dependent power law density profile
for all our models presented here,
\begin {equation}
n_{H}(r)=n_{H}(r_{0})\times (r/r_{0})^{\alpha}.     
\end {equation}
Here $n_{H}(r)$, $n_{H}(r_{0})$ are the density at a distance $r$ and at the illuminated face at $r_{0}$, respectively. The power law index $\alpha$
can be varied. Previous modellers \citep{2005Boirin} used the simplest plane 
parallel geometry. 
In a plane parallel geometry, the ratio of thickness to the inner radius (r-r$_0$)/r$_0$ is < 0.1. Hence, for a plane parallel model, the value of power 
law index does not affect the 
results. However, for a thick shell model, where (r-r$_0$)/r$_0$ is < 3), the results do depend on the value of power law index. Here, we consider 
a more realistic thick shell geometry and we use a power law index of $-15/8$ \citep{Frank2002}.

\citet{1997Gingamemsympo} showed that the X-ray spectrum of LMXBs can be of following 4 types : i) hard power law with a sharp cut-off 
above a few tens of keV, ii) soft thermal component accompanied by a hard tail, iii) thermal bremsstrahlung, and iv) a single power law. 
Here we assume that the gaseous disc is exposed to incident radiation coming 
from a thermal blackbody and a single power law due to Comptonisation, following \citet{Juett2005} and B05. We use ionisation parameter 
$\xi$ (erg cm s$^{-1}$) to quantify this radiation, 
and it is given by,
\begin {equation}
 \xi = \frac{L_{ion}}{n(H)r^{2}} = (4\pi)^{2}\int_{1R}^{1000R}{\frac{J_{\nu}d\nu}{n(H)}} . 
\end {equation}
Here $J_{\nu}$ is the mean intensity in erg s$^{-1}$ sr$^{-1}$ Hz$^{-1}$. The integration limit ranges from $R$ to 1000 $R$ in units of Ry, Ry being the ionization 
energy of hydrogen atom. Following 
\citet{2001ApJS...133...221}, we include all ionising radiation, integrating over all ionising photon energies from $1R$ to $1000R$. 
The source--gaseous-disc separation is represented by $r$. 
$L_{ion}$ is the luminosity between 1 and $10^{3}$ Ry. 

\citet{2005Boirin} and earlier modellers assumed plane parallel geometry without any wind. We find that a rotating disc model produces smaller $\chi^{2}$ than a plane parallel geometry, and hence, we assume a rotating disc with wind in all our calculations presented here.  
For a rotating disc geometry, 
the inward gravitational acceleration is calculated as
\begin {equation}
g = \frac{GM}{r^{2}}(1-\frac{r_{0}}{r})\,,      
\end {equation}
where $r_{0}$ is the inner radius. In this given set up, CLOUDY evaluates the line widths and escape probabilities of photons in the Sobolev or Large 
Velocity Gradient approximation \citep{1957SvA...1...678S}. The effective line optical depth is given by
\begin {equation}
\tau_{l,r}=\alpha_{l,u}min(r,\Delta r)(n_{l}-n_{u}\frac{g_{l}}{g_{u}})(\frac{u_{th}}{max(u_{th},u_{exp})}).    
\end {equation}
Here, $n_{l}$, $n_{u}$, $g_{l}$, $g_{u}$ and $\alpha_{l,u}$ are level populations and statistical weights 
in lower and upper levels and line absorption co-efficient, respectively. $u_{th}$ and $u_{exp}$ are the thermal and expansion velocities respectively,
and the radius used is the
smaller of the depth ($\Delta r$) or the radius ($r$).

Other modellers, whom we mentioned earlier, ignored the magnetic field in their calculations. However, as we mentioned in the introduction section 
that the effect of magnetic fields can be important in accretion discs, we include the magnetic field in our models as one of the input parameters, and study 
its
effect on highly ionised soft x-ray lines. Here, as mentioned below, we suggest a plausible way to estimate the upper limit of LMXB disc magnetic field.
It is known that the 
magnetic field cools high-temperature ionised gas through cyclotron emission and takes 
dominant part in thermal cooling and contributes to gas pressure. Hence this field is capable to change the strength and profile of emission or 
absorption lines from a high-temperature ionised gas. Therefore, the strengths and profiles of lines originating in accretion discs 
can be utilised to estimate an upper limit of the local magnetic field. Such a field prevailing in the accretion discs can be 
either ordered or tangled. We do not include ordered magnetic field in our calculation as the angle between the radiation 
field of the central object and the magnetic field is unknown. Our used tangled magnetic field (in units of G) follows the equation
\begin {equation}
B = B_{0} \times (n_{H}(r) / n_{H}(r_{0}))^{\gamma/2}.
\end {equation}
Here, $B_{0}$ is a free parameter and  the term in parenthesis is the ratio of the  density at a distance $r$ to the density at the 
illuminated face of the gaseous disc at $r_{0}$. $\gamma$ depends on geometry and for the tangled field we use the 
default value provided by CLOUDY, $\gamma$ = 4/3. 

All the models calculated here are constant pressure models. The total pressure consists of ram, magnetic, turbulent, particle, and
radiation pressures. The magnetic pressure and the enthalpy terms are also considered in these models.

CLOUDY requires at least one  stopping condition to complete its calculation. Here, we use ionised column density as one of the stopping conditions 
for our calculations. No observed ionised column density is reported yet for the above mentioned systems, so we vary the ionised column density 
in our calculations. Our other stopping condition is the thickness of the accretion disc. CLOUDY stops calculation as soon as it satisfies any one of 
these two conditions mentioned. 
Best model is calculated by using a built-in optimisation program based on phymir algorithm \citep{vanhoof1997} which calculates a non-standard 
goodness-of-fit estimator $\chi^2$ and minimises it by varying input parameters. This optimisation program is user friendly and preferred by many 
astronomers \citep{{Ferland2013}, {2000Srianand}, {vanhoof2013},{2018Rawlins}}. We try to get best model parameters by varying total hydrogen density, 
ionisation parameters, metallicity, clumpiness, magnetic field and ionised column density. CLOUDY executes detailed
radiative transfer computations and produces line fluxes, column densities of various species, heating and cooling, 
self-consistently. In the following, we show the effect of important parameters on the Fe XXV and Fe XXVI lines, since these lines are present in all the four LMXBs modelled here. Other highly ionised lines mentioned in this work also show similar effects with different extents.

\subsection{Effect of density} \label{subsec:density}
\begin {figure}
%  \centering
  \vspace*{-0.2in}
  \hspace*{-0.1in}
\includegraphics[scale=0.48]{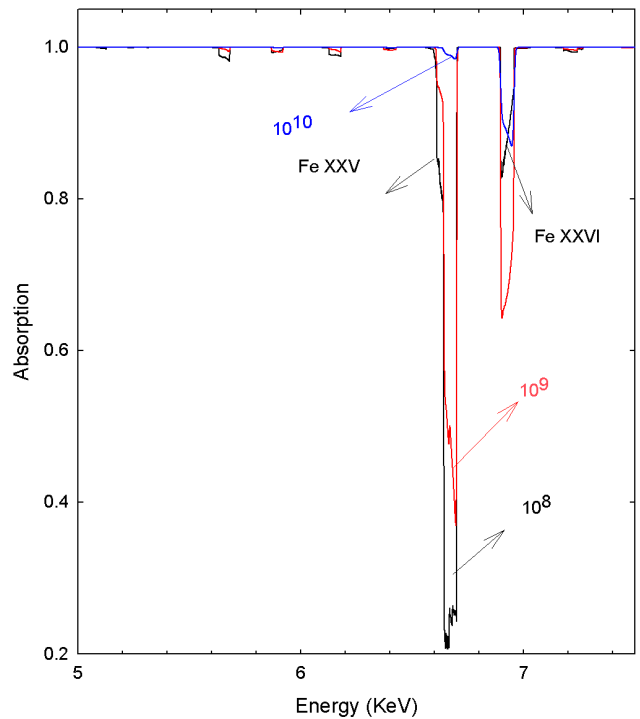}
%\plotone{fig1.pdf}
\vspace*{-1.2in}
\caption{Change in relative absorption strength of Fe XXV and Fe XXVI 
lines with changing density are plotted as a function of energy in the energy range $5 - 7.5$ keV. Here black, red and blue solid lines represent models with 
densities 10$^{8}$ cm$^{-3}$, 10$^{9}$ cm$^{-3}$, 10$^{10}$ cm$^{-3}$ at the illuminated face of the accretion disc, respectively (see section 2.2).\label{fig:f1}}
\end {figure}

First, we consider a sample model for LMXBs where we assume a geometry of a constant pressure clumpy highly-ionised gaseous rotating disc around a 1.4 solar 
mass neutron star with wind velocity 700 km s$^{-1}$. The sample model parameters are shown in Table \ref{tab:table 1}. The effects of various parameters are shown by varying 
them one at a time, keeping others fixed. We show the effect of density parameter on  Fe XXV and 
Fe XXVI lines by varying the density parameter alone keeping all other parameters fixed. 
Fig. \ref{fig:f1} shows the absorption strength of Fe XXV and Fe XXVI lines as a function of energy for three different 
densities at the illuminated face of the accretion disc, 10$^{10}$, 10$^{9}$ and 10$^{8}$ cm$^{-3}$, represented by blue, red and black solid lines, 
respectively. 
Absorption strengths of both Fe XXV and Fe XXVI 
lines change by a change in density. However in this case, a change in density does not produce any change in line width. 
For this sample model, above a particular density at the illuminated face of the accretion disc ($\sim$ 10$^{10}$ cm$^{-3}$), the strength of Fe XXV line 
is much less than that of 
Fe XXVI. Below this density, the strength of Fe XXV is higher. This may be due to critical density effect. An increase in density at the illuminated face of the accretion disc from 10$^{8}$ cm$^{-3}$ to 10$^{9}$ cm$^{-3}$ 
increases the absorption for 
Fe XXVI line roughly by 100 \% and decrease in absorption for line Fe XXV roughly by 33\%. Besides Fe XXV and Fe XXVI lines,
very weak highly ionised lines from other species 
can  also be found in this simple sample model.

\begin{table}
	\centering
	\caption{Input parameters for our sample LMXB model.}
	\label{tab:table 1}
	\begin{tabular}{lr} 
		\hline
		Physical parameters & Values\\
		\hline
%		\begin{deluxetable}{ccc}
%\tablenum{1}
%\tablecaption{Predicted physical parameters by CLOUDY modelling of 4U 1323--62~\label{tab:tab1}.}
%\tablewidth{0pt}
%\tablehead{
%\colhead{Physical parameters} & \colhead{Persistent} & \colhead{Dip}
%}
%\decimalcolnumbers
%\startdata
Power law: slope, log($\xi$) & -1.96, 3.8 \\
Blackbody: Temperature, log($\xi$) & 10$^{7}$ K, 3.2 \\
Density n(r$_{0}$) (cm$^{-3}$) & 10$^{9.5}$ \\
Metallicity & solar \\
Clump filling & 0.7  \\
Wind (km s$^{-1}$) & 700  \\
Ionised column density (cm$^{-2}$) & 22.8 \\
%\enddata
%\end{deluxetable} 
\hline
\end{tabular}
\end{table}

\subsection{Effect of radiation field} \label{subsec:radiation}
\begin {figure}
% \centreing
  \vspace*{-0.1in}
  \hspace*{-0.1in}
\includegraphics[scale=0.48]{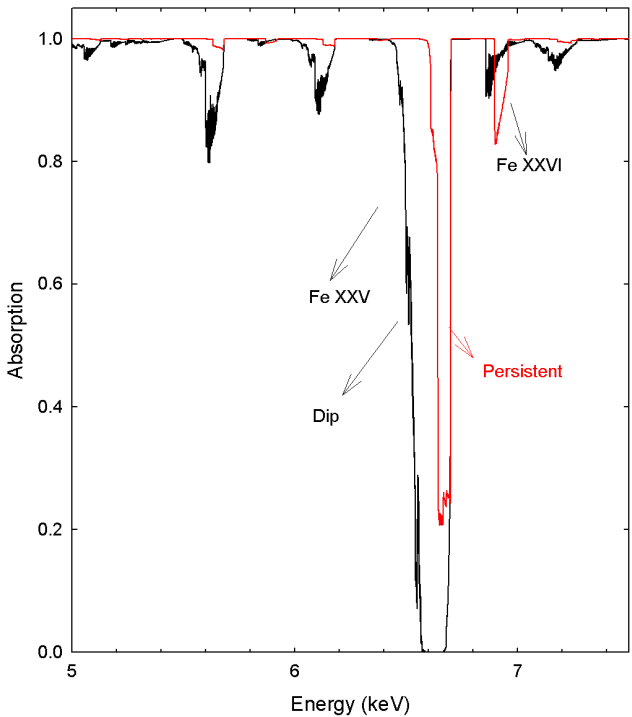}
%\plotone{fig1.pdf}
\vspace*{-1.2in}
\caption{Change in relative absorption strength of Fe XXV and Fe XXVI lines are plotted as a function of energy in the energy range $5 - 7.5$ keV.
The red and black solid lines represent cases with higher and lower ionisation parameter, respectively (see section 2.3). 
\label{fig:f2}}
\end {figure}

In all our models, we consider incident radiation to be consisted of both power-law continuum and blackbody continuum. 
In Fig. \ref{fig:f2} we plot absorption strengths of Fe XXV and Fe XXVI lines as a function of energy with different 
ionisation parameters for the same sample model of LMXB (see Table \ref{tab:table 1}). Here we vary the ionisation parameter keeping all other parameters fixed.
A decrease in ionisation parameter increases the line flux and line width of Fe XXV. A decrease in ionisation parameter from 3.8 to 3.3 increases the Fe XXV 
line absorption five times at the line centre. It has been observed by B05 and \citet{Trigo2006}
that in the dipping phase of LMXB dippers, line flux and line width of Fe XXV 
increases. So Fig. \ref{fig:f2} can also demonstrate that ionisation parameter 
is lower for the dipping phase compared to the persistent phase of LMXBs.

\subsection{Effect of wind} \label{subsec:wind}
\begin {figure}
% \centering
  \vspace*{-.3in}
  \hspace*{-.2in}
\includegraphics[scale=0.48]{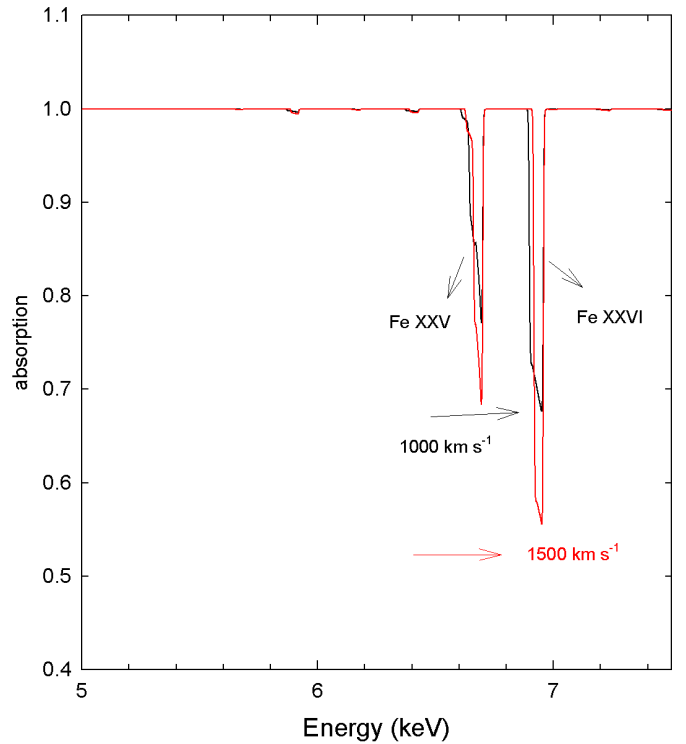}
%\plotone{fig1.pdf}
\vspace*{-0.9in}
\caption{Change in relative absorption strength of Fe XXV and Fe XXVI lines are plotted as a function of energy in the energy range $5 - 7.5$ keV for disc wind 
velocities 
1000 km s$^{-1}$ (black solid line) and 1500 km s$^{-1}$ (red solid line) respectively (see section 2.4). 
\label{fig:f3}}
\end {figure}

Earlier modellers like B05, \citet{Trigo2006} and \citet {Juett2005} did not include disc wind in their models.
To make our model more realistic \citep{D'Ai2014}, we include disc wind and it is one of the important input parameters. Fig. \ref{fig:f3} shows the effect of different wind velocities 
on the absorption strength of Fe XXV and Fe XXVI lines for the same sample model discussed earlier (see Table \ref{tab:table 1}). Here we keep all the parameters of 
the sample model fixed and vary only the wind velocity. An increase in wind velocity from 1000 km s$^{-1}$ to 1500 km s$^{-1}$ increases both the line strengths 
by $\approx$ 20$\%$. In addition, 
metallicity and the clumpiness also effects the strength of the iron line absorption but to a lesser extent. 

\subsection{Effect of magnetic field} \label{subsec:field}
As we mentioned previously, in the presence of magnetic fields these absorption lines can get modified. Through the amount of change, one can estimate an upper 
limit on 
the strength of magnetic field in these accretion discs. Here we demonstrate that by comparing models with and without magnetic field.

\begin {figure}
% \centreing
  \vspace*{-0.7in}
  \hspace*{-0.55in}
\includegraphics[scale=0.48]{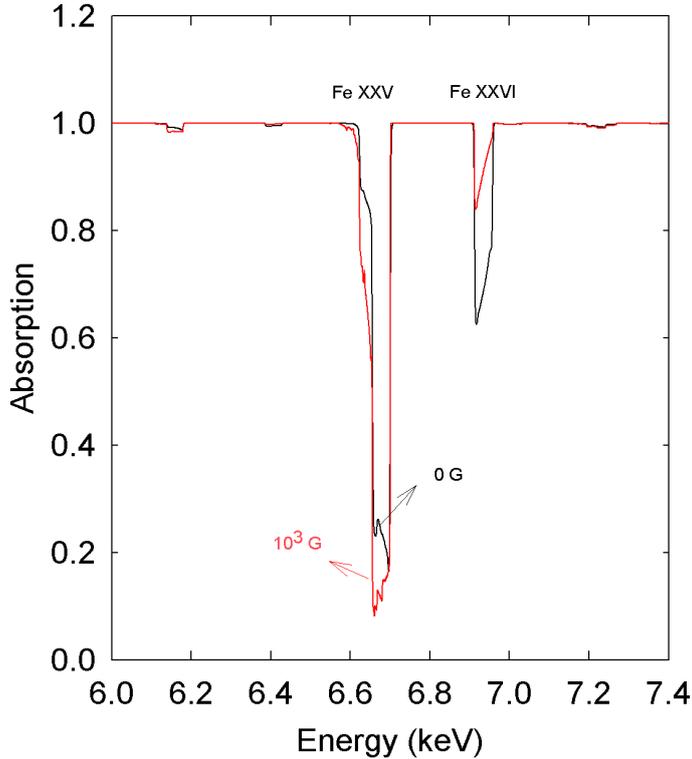}
%\plotone{fig3.pdf}
\vspace*{-0.5in}
\caption{Effect of magnetic field on Fe XXV and Fe XXVI lines are plotted as a 
function of energy in the energy range 6- 7.4 keV. 
The solid black and red lines represent models with no-magnetic field and $10^{3}$ Gauss magnetic field, respectively (see section 2.5).}
\label{fig:f4}
\end {figure}

Fig. \ref{fig:f4} shows the effects of  magnetic field strengths on Fe XXV and Fe XXVI lines.
Here we represent models with no-magnetic field and $10^{3}$ Gauss magnetic field by the solid black and red lines, respectively.  An increase in magnetic field to 10$^{3}$ G from null increases 
the line intensity of Fe XXV line approximately by 10\% and decreases the line intensity of Fe XXVI line by 32\%.
Higher magnetic field produces more significant effect. Thus, an upper limit of magnetic field can be
derived using the observed line fluxes of Fe XXV and Fe XXVI. The detectable amount of change in the 
absorption due to a few hundred Gauss magnetic field is the effect of change in temperature and pressure.
A high magnetic field cools ionised gas through cyclotron emission and it is the dominant coolant. Our calculation shows that the gas 
temperature in the accretion disc decreases nearly by an order or factor of ten for the model with $10^{3}$ Gauss magnetic field compared to the zero magnetic field model.

\section{Results}
In this section we present our results and elaborate on main findings. First we  discuss results of GX 13+1 in detail as it has the maximum number of 
observed lines. Then we briefly discuss the results for MXB 1659--298, 4U 1323--62 and 4U 1916--053.
\subsection{GX 13+1} \label{subsec:GX 13+1}
\begin{figure}
	% To include a figure from a file named example.*
	% Allowable file formats are eps or ps if compiling using latex
	% or pdf, png, jpg if compiling using pdflatex	
%\plotone{fig1.pdf}
  %\vspace{-0.22in}
  \hspace*{-0.2in}
  \includegraphics[scale=0.5]{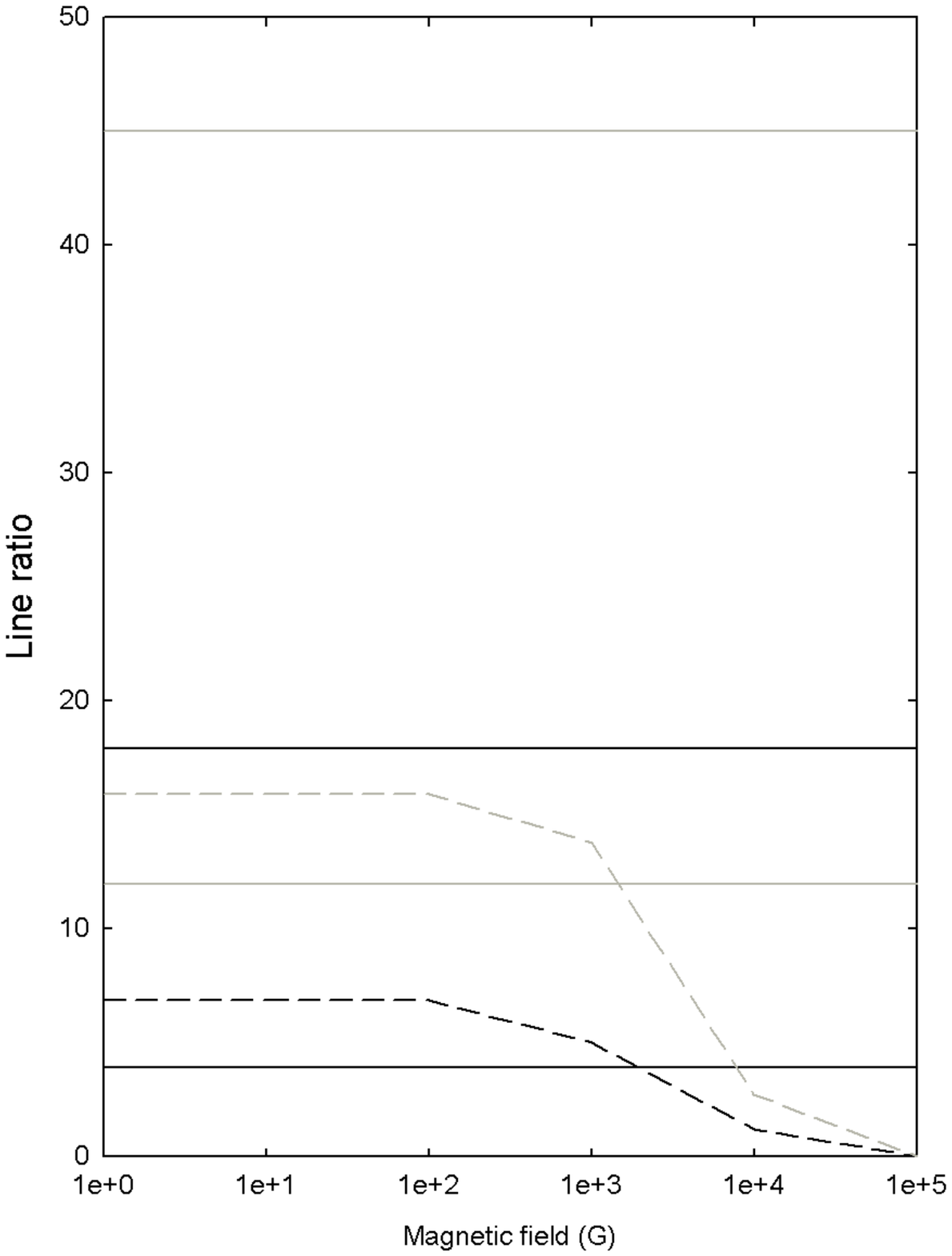}
\vspace*{-.5in}
    \caption{Predicted Fe XXV line ratio for GX 13+1 is plotted as a function of magnetic field. The horizontal lines represent the observed range (see section 3.1).
    The black and gray lines represent data for Fe XXV and Fe XXVI, respectively.
    }
    \label{fig:f5} 
\end{figure}

\begin{figure}
	% To include a figure from a file named example.*
	% Allowable file formats are eps or ps if compiling using latex
	% or pdf, png, jpg if compiling using pdflatex
	
%\plotone{fig1.pdf}
  %\vspace*{-0.25in}
  \hspace*{-0.15in}
  \includegraphics[width=\columnwidth]{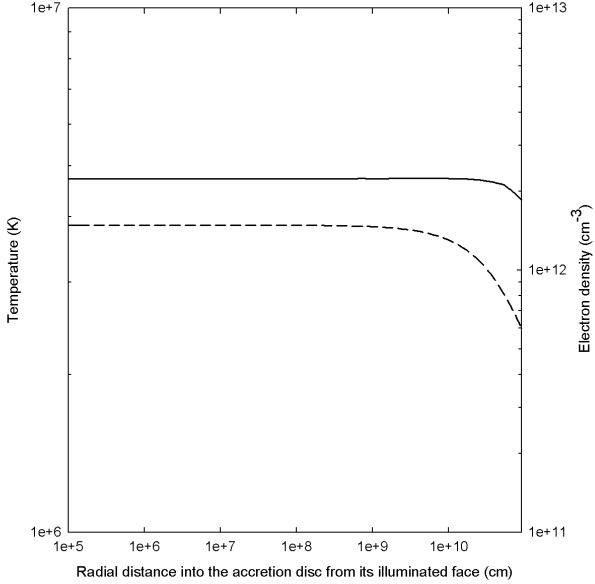}
    \caption{Temperature and electron density are plotted as a function of distance into the accretion disc. The solid and dashed lines represent temperature 
    and electron density, respectively (see section 3.1.)}
    \label{fig:f6} 
\end{figure}

\begin{figure}
	% To include a figure from a file named example.*
	% Allowable file formats are eps or ps if compiling using latex
	% or pdf, png, jpg if compiling using pdflatex
	
%\includegraphics[scale=0.4]{fig3.pdf}
%\plotone{fig1.pdf}
	\includegraphics[width=\columnwidth]{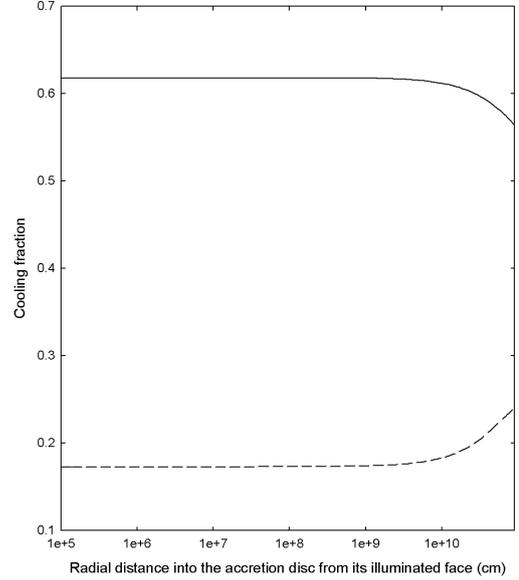}
    \caption{Cooling fractions for important coolants are plotted as a function of distance into the accretion disc. The solid and medium-dashed lines 
    represent cooling by free-free transition and adiabatic cooling, respectively (see section 3.1).}
    \label{fig:f7}
\end{figure}

\begin{figure}
	% To include a figure from a file named example.*
	% Allowable file formats are eps or ps if compiling using latex
	% or pdf, png, jpg if compiling using pdflatex
	
%\includegraphics[scale=0.4]{gx13p_fig}
%\plotone{fig1.pdf}
	\includegraphics[width=\columnwidth]{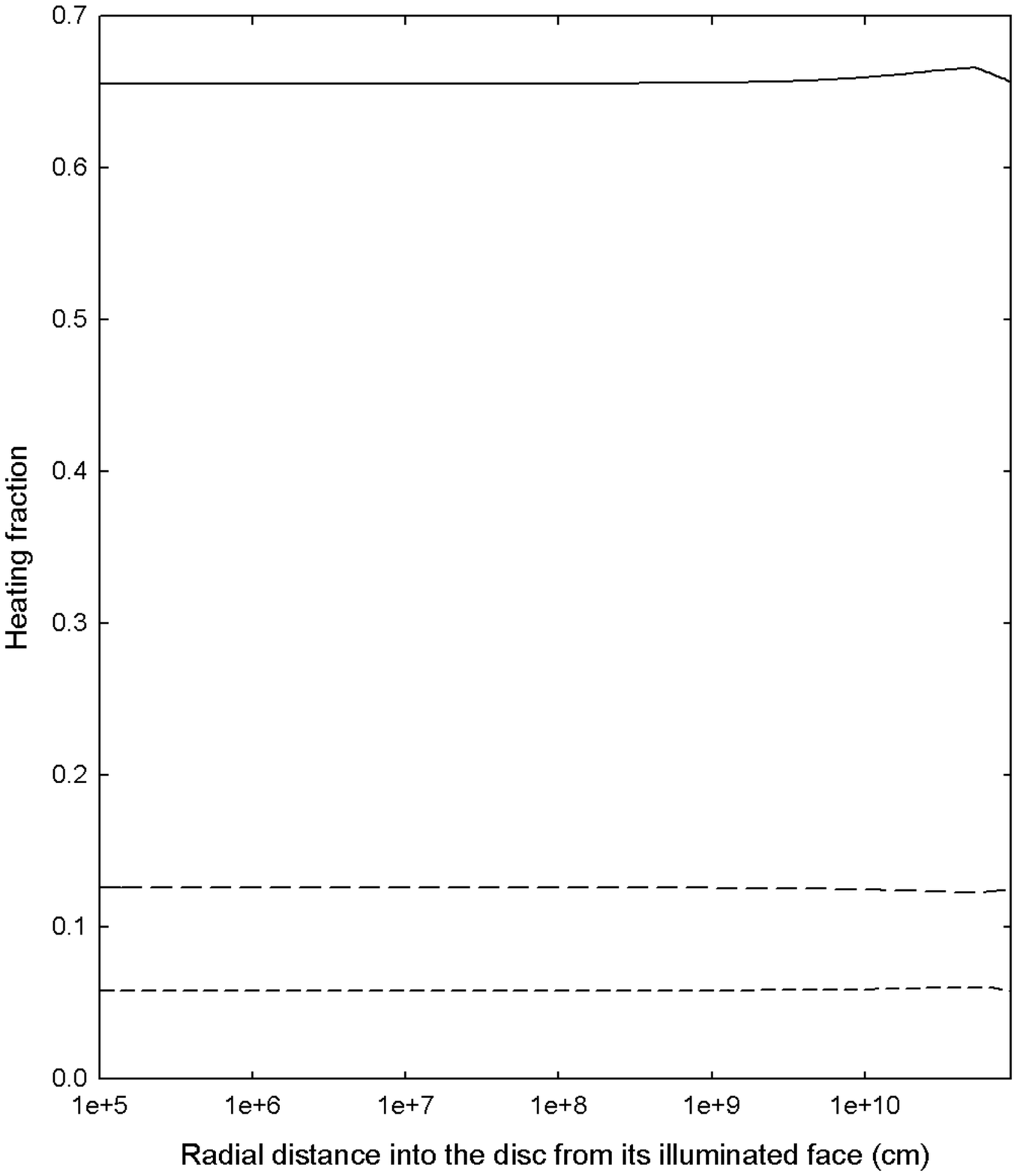}
    \caption{Heating fractions for important heating sources are plotted as a function of distance into the gaseous disc. The solid, 
    medium-dashed and short-dashed lines represent Compton heating and heating by species Fe25 and Fe24 respectively (see section 3.1.)}
    \label{fig:f8}
\end{figure}

Here we model the persistent phase of GX 13+1 reported by \citet{D'Ai2014}. They  observed absorption lines Fe XXV, Fe XXVI, Mg XII, Al XIII, 
Si XIV, S XVI, Ca XX and listed  the line fluxes. We normalise line fluxes with respect to the weakest Al XIII line flux. \citet{D'Ai2014} have measured a common blue-shift (weighted
average 490 km s$^{-1}$ ) for resonant transitions of H-like ions coming from the accretion disc of GX 13+1. We fix the wind velocity to  
500 km s$^{-1}$ (close to the observed value) for model calculations presented here for GX 13+1.
The built-in CLOUDY optimisation program (see end of subsection 2.1) is widely used by 
astronomers \citep{{Ferland2013}, {2000Srianand}, {vanhoof2013},{2018Rawlins}} to match model predictions close to observed values. 
Applying the same approach, the optimisation models as mentioned earlier with different power law index values are performed, and we notice that the model with power 
index value -0.6 gives the best result.
The gas phase abundances of the observed elements Fe, Mg, Ca, S and Si are varied to match with observations and  
abundances of other elements are fixed to their solar values due to lack of observed lines. Dusts are not considered in this model as the  
temperature is much higher than the sublimation temperature of dusts. First, we calculate a best model without any magnetic field. In the 
next step, we add a tangled magnetic field in our model and vary its strength. The predicted line ratios start going beyond the observed range above 2000 G. 
The most affected lines are highly ionised Fe lines and the least affected lines are highly ionised Mg lines. As an example, we plot our model predictions as a
function of magnetic field for lines Fe XXV and Fe XXVI in Fig.\ref{fig:f5}. The predicted Fe XXV and Fe XXVI line ratios remain within the observed 
range for magnetic field less than 
10$^3$ G. We also show  results for all the detected lines in Table \ref{tab:table 2} for various magnetic field values. From our calculations, 
we can estimate a magnetic field upper limit 
of close to 2000 G. Note that our results do show (see Table \ref{tab:table 2} and Fig.\ref{fig:f5} ) that for a much higher magnetic field value, the predicted line ratios are far from 
the observed ranges. Normalisation with respect to a different line change the line ratios. But the overall physical conditions and the 
upper limit of magnetic field remain almost the same as the choice of normalisation does not change the system. Table \ref{tab:table 3} lists predicted physical parameters for our best model. Our predicted hydrogen 
density is $10^{12.1}$ cm$^{-3}$ 
which is close to that predicted by \citet{D'Ai2014} who reported a density of $10^{12}$ cm$^{-3}$. As mentioned earlier, we consider incident radiation field 
consisting of a power law continuum and a blackbody radiation. Earlier, \citet{Migliari2005} found the power index  to be -0.6. Whereas, \citet{Paizis2006} 
have calculated a power index of $-2.8^{+1.48}_{-1.33}$. We try three power indexes $-0.6, -1$ and $-2$. For our best 
model, the power law  radiation has a spectral index $-0.6$ and the temperature of the blackbody radiation is $3\times 10^{7}$ K. 
Similar temperature for 
the blackbody radiation has been reported by others too. Our best ionisation parameters for these two radiations are 3.0 and 1.0 (as mentioned earlier), 
respectively. In general, radial thickness of the accretion disc from its illuminated face ranges from 10$^{9}$ cm to $10^{11}$ cm. Here, 
the best model predicts that the above mentioned lines originate near the edge of the disc, close to $1.6\times
10^{11}$cm, and the radial thickness of the accretion disc from its illuminated face to be $10^{11}$ cm. \citet{D'Ai2014} had mentioned a 
similar value of  $1.4\times 10^{11}$ cm. Accretion disc is highly ionised and our predicted ionised column density is $10^{23.0}$ cm$^{-2}$.  
Elemental abundances differ from their solar values with different amounts for this LMXB. Elemental abundances of Ca and Si are close to their respective 
solar values,
whereas elemental abundance of Fe is more than its respective  solar value. Table \ref{tab:table 4} compares the predicted and observed 
line flux ratios with respect to the weakest line Al XIII for our best model. All the predicted line flux ratios are within the observed range. 
Besides the lines mentioned here, our best model predicts Ar XVIII and Ni XXVIII lines at 3.735 {\AA} and 1.533 {\AA}, respectively. 
These two lines are not reported by the observers. In our calculation we consider solar abundances for Ar and Ni. 
Lack of observation despite model prediction might mean that the abundances of Ar and Ni are less than their respective solar values. 
As mentioned earlier, we include a tangled magnetic field in our calculation. Comparing our model predictions and observed values, we predict
an upper limit of magnetic field of strength $\approx$ 1000 G in the accretion disc. Fig.\ref{fig:f6} shows the ranges of temperature and electron density across the accretion disc as a function of depth from its illuminated face.  
The solid and dashed lines represent temperature and electron density, respectively. The temperature in the disc ranges from $4.72\times 10^{6}$K to $4.24\times 10^{6}$K,
whereas the electron density ranges from $1.48\times 10^{12}$ cm$^{-3}$ to $5.6\times 10^{11}$ cm$^{-3}$.
In CLOUDY, temperature is determined by solving heating and cooling balance. It includes all the important heating and cooling mechanisms 
self-consistently to determine the temperature accurately. In Fig.\ref{fig:f7} we plot the cooling fraction of important coolants across the accretion disc 
as a function of depth from the illuminated face of the accretion disc. Most of the cooling, roughly 60 percent is contributed by free-free 
cooling. 
Rest are coming from adiabatic  cooling and Compton cooling, respectively.  
In Fig.\ref{fig:f8}, we show the heating fraction as a function of depth from the illuminated face of the accretion disc. Most of the heating comes 
from Compton heating, which is again roughly 65 percent. The other important heating agents are heating by species Fe25 and Fe26, respectively.

\begin{table}
	\centering
	\caption{Comparison of observed and predicted line flux ratios  of LMXB GX 13+1 with different magnetic field strengths (in units of G) using CLOUDY. 
	Lines are normalised w.r.t Al XIII line.}
	\label{tab:table 2}
	\begin{tabular}{llllll} 
		\hline
		Lines & \hspace*{-0.3in} Observed & \hspace*{-0.3in} Model& \hspace*{-0.5in} Model& \hspace*{-0.7in} Model& \hspace*{-0.3in} Model\\
		&\hspace*{-0.15in} range & \hspace*{-0.3in} B=0 & \hspace*{-0.5in} B=10$^2$ & \hspace*{-0.7in} B=10$^4$& \hspace*{-0.3in} B=10$^5$\\ 
		\hline
		Mg XII$^a$ &\hspace*{-0.2in}0.39--3.23 &\hspace*{-0.2in}2.96 &\hspace*{-0.4in}2.95&\hspace*{-0.6in}3.47&\hspace*{-0.2in}5.16\\
		Mg XII$^b$& \hspace*{-0.2in}0.54--3.22 & \hspace*{-0.2in}0.88&\hspace*{-0.4in}0.88&\hspace*{-0.6in}0.56&\hspace*{-0.2in}0.27\\
		Si XIV$^c$&\hspace*{-0.2in}2.02--7.47 & \hspace*{-0.2in}4.99&\hspace*{-0.4in}4.98&\hspace*{-0.6in}6.53&\hspace*{-0.2in}6.16\\
		S XVI$^d$& \hspace*{-0.2in}0.93--4.48 & \hspace*{-0.2in}1.86&\hspace*{-0.4in}1.86&\hspace*{-0.6in}1.41&\hspace*{-0.2in}0.79\\
		Ca XX$^e$& \hspace*{-0.2in}1.12--8.14 & \hspace*{-0.2in}2.49&\hspace*{-0.4in}2.49&\hspace*{-0.6in}1.27&\hspace*{-0.2in}0.02\\
		Fe XXV$^f$& \hspace*{-0.2in}3.91--17.92 & \hspace*{-0.2in}6.83&\hspace*{-0.4in}6.83&\hspace*{-0.6in}1.19&\hspace*{-0.2in}<0.005\\
		Fe XXVI$^g$& \hspace*{-0.2in}11.95--45.0 &\hspace*{-0.2in}15.92&\hspace*{-0.4in}15.89&\hspace*{-0.6in}2.71&\hspace*{-0.2in}<0.005\\
		Fe XXVI$^h$& \hspace*{-0.2in}3.16--28 & \hspace*{-0.2in}3.20&\hspace*{-0.4in}3.20&\hspace*{-0.6in}0.59&\hspace*{-0.2in}<0.005\\
		\hline
$^a$1.472 keV, & $^b$1.744 keV, & $^c$2.005 keV, & $^d$2.621 keV,& &\\
$^e$4.105 keV, &$^f$6.700 keV, &$^g$6.966 keV, & $^h$8.250 keV. &&\\
%\multicolumn{1}{l}{$^a$1.472 keV},
%\multicolumn{3}{l}{$^b$1.744 keV},
%\multicolumn{3}{l}{$^c$2.005 keV}, 
%\multicolumn{1}{l}{$^d$2.621 keV}, 
%\multicolumn{2}{l}{$^e$4.105 keV},
%\multicolumn{2}{l}{$^f$6.700 keV},
%\multicolumn{2}{l}{$^g$6.966 keV},
%\multicolumn{3}{l}{$^h$8.250 keV}\\
%				
	\end{tabular}
\end{table}

% Example table
\begin{table}
	\centering
	\caption{Predicted model parameters of LMXB GX 13+1 using CLOUDY.}
	\label{tab:table 3}
	\begin{tabular}{lr} 
		\hline
		Physical parameters & Best values\\
		\hline
		Power law: slope, log($\xi$) & -0.6, 3.0\\
		Blackbody: Temperature, log($\xi$) & $3\times 10^{7}$ K, 1 \\
		Density n(H) ($cm^{-3}$) & $10^{12.1}$\\
		Ionised column density ($cm^{-2}$) & 23.2\\
		Upper limit of magnetic field (G) & 2000\\
		Fe/H & -4.31 \\
		Mg/H & -4.64 \\
		Ca/H & -5.7 \\
		S/H & -5.56 \\
		Si/H & -4.5 \\
		\hline
				
	\end{tabular}
\end{table}

\begin{table}
	\centering
	\caption{Comparison of observed and predicted best line flux ratios  of LMXB GX 13+1 using CLOUDY. Lines are normalised w.r.t Al XIII line. }
	\label{tab:table 4}
	\begin{tabular}{lcr} 
		\hline
		Lines $E_{lab}$(kev)  & Observed & Best model \\ 
		\hline
		Mg XII (1.472) &0.39--3.23 & 2.96\\
		Mg XII (1.744)& 0.54--3.22 & 0.88 \\
		Si XIV (2.005)& 2.02--7.47 &  4.99\\
		S XVI (2.621)& 0.93--4.48 & 1.86\\
		Ca XX (4.105)& 1.117--8.139 & 2.49\\
		Fe XXV (6.700)& 3.91--17.92 & 6.83\\
		Fe XXVI (6.966)& 11.95--45.0 & 15.92\\
		Fe XXVI (8.250)& 3.16--28 & 3.2\\
		\hline
				
	\end{tabular}
\end{table}
% Example figure

\label{sec:Results}
\subsection{MXB 1659--298} \label{subsec:MXB 1659--298}
MXB 1659--298 is also a NS-LMXB which shows type-I X-ray
bursts. Recently \citet{2018Iariab} have observed O VIII, Ne
X, Fe XXV, Fe XXVI, Fe XXV and Fe XXVI lines for MXB 1659--298 using XMM-Newton,
$\it NuSTAR$ and $\it Swift/XRT$. We follow a similar model for MXB 1659--298 like the one used for 
GX 13+1. We consider the observed line fluxes reported by
\citet{2018Iariab} and use CLOUDY to model these lines to understand
the underlying physical conditions. In this case, we normalise line fluxes with respect to the weakest O
VIII line flux. Unlike GX 13+1, the wind velocity is not
detected unambiguously here. So, we keep wind velocity as a free
parameter for the model of MXB 1659--298. Tabel \ref{tab:table 5} and Table \ref{tab:table 6} list the
best model parameters and predicted line flux ratios using CLOUDY.
Our predicted hydrogen density is 10$^{10.45}$ cm$^{-3}$. For our
best model, the power law radiation has a spectral index of $-2.1$
and the temperature of the blackbody radiation is $5\times
10^{7}$K. The ionisation parameters for power-law and black
body radiations are 2.94 and 1.66, respectively. Here, the best model predicts the radial expansion
of the accretion disc from its illuminated face to be $1\times 10^{11}$ cm, whereas the predicted
ionised column density is 10$^{23}$ cm$^{-2}$. Elemental abundances of
Fe, O and Ne are close to their solar values. All the predicted line flux
ratios except the Fe XXVI (8.250keV) are within the observed
range. Prediction for the Fe XXVI line does not
improve even when we include all the iso-electronic levels of
Fe. However,
for completeness we show our prediction for this
line in Table \ref{tab:table 6}. As mentioned earlier, in the final stage
we include a tangled magnetic field in our calculation. We find that, with the upper limit magnetic field $\approx$ 1000 G, all
the Fe line ratios exceed the observed ranges. No such drastic effect is
observed for Ne X line.
\begin{table}
	\centering
	\caption{Predicted model parameters of MXB 1659--298 using CLOUDY.}
	\label{tab:table 5}
	\begin{tabular}{lr} 
		\hline
		Physical parameters & Best values\\
		\hline
		Power law: slope, log($\xi$) & -2.1, 2.94\\
		Blackbody: Temperature, log($\xi$) & $5\times 10^{7}$ K, 1.66 \\
		Density n(H) ($cm^{-3}$) & $10^{10.8}$\\
		Clump filling & 0.6\\
		Wind (km/sec) & 700\\
		Ionised column density ($cm^{-2}$) & 23.0\\
		Upper limit of Magnetic field (G) & $\approx$ 1000 \\
		Fe/H & -4.4 \\
		O/H & -4.32 \\
		Ne/H & -4.1 \\
		\hline		
	\end{tabular}
\end{table}

\begin{table}
	\centering
	\caption{Comparison of observed and predicted line flux ratios  of MXB 1659--298 using CLOUDY. Lines are normalised w.r.t O VIII line.}
	\label{tab:table 6}
	\begin{tabular}{lcr} 
		\hline
		Lines $E_{lab}$(kev)  & Observed & Predicted\\
		\hline
		Ne X (1.021)&6.09--3.01  & 4.98\\
		Fe XXV (6.700)&28.7--16.10 &27.60\\
		Fe XXVI (6.966)&36.23--21  & 23.61\\
		Fe XXV (7.880)& 27.72--9.47 & 10.64\\
		Fe XXVI (8.250)& 30.29--12.62 & 4.05\\
		\hline
				
	\end{tabular}
\end{table}

\label{sec:Results}

\subsection{4U 1323--62} \label{subsec:4U 1323--62}

\label{sec:Results} % used for referring to this section from elsewhere

Here, we model the spectral lines of 4U 1323--62. In this case, observed absorption fluxes of Fe XXV and Fe XXVI lines are taken from the panel A of 
Figure 2 of B05, and we try to 
match the flux ratios of these two lines (Fe XXV/Fe XXVI) at the line centre. 
We vary all the parameters of our sample LMXB dipper model as discussed earlier and try to match the predicted absorption ratio of Fe XXV and Fe XXVI lines 
with the observed one. In Table \ref{tab:table 7}, list of predicted parameters for 
the best model 
of 4U 1323--62 are shown. Our predicted ionisation parameters and ionised column densities in the persistent and dipping phases are very close to that 
predicted by B05. The best model 
predicts that the metallicity is sub-solar and clumps are present in the gaseous disc. This sub-solar metallicity may indicate 
the metallicity of the donor star. Earlier, \citet{{Zolotukhin2010},{Gambino2016}} had estimated the mass of the donor star to be 
0.28$\pm$0.03 times solar mass but did not discuss about its metallicity. We predict that wind velocity, density and metallicity remain to be
almost the same for both the persistent and dip phases, whereas the ionisation 
parameter and the ionised column density are the only parameters that differ in persistent and dip phases. Earlier, B05 used 
more than one incident radiation fields in their modelling. We also predict an incident radiation field consisting of a blackbody radiation of temperature 
$10^{7}$K and Comptonised radiation with power law index $-1.99$. 
Here wind velocity is 
700 km s$^{-1}$. Note that
\citet{2005ApJ...620...274} also observed similar wind velocities for Galactic LMXBs earlier. 
 
\begin {figure}
% \centreing
  \vspace*{-0.2in}
  \hspace*{-0.1in}
\includegraphics[scale=0.48]{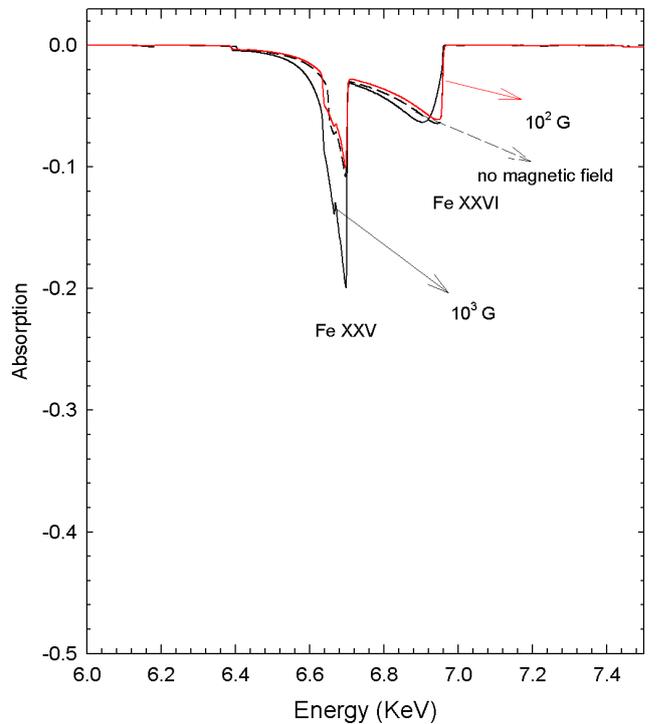}
%\plotone{fig3.pdf}
\vspace*{-1.2in}
\caption{Effect of magnetic field on the modelled spectra for 4U 1323--62 in persistent phase is plotted as a 
function of energy in the energy range 6 - 7.5 keV. 
The  medium-dashed black line, solid red line and solid black line represents 
models with no-magnetic field, $10^{2}$ and $10^{3}$ Gauss magnetic field, respectively (see section 3.3). 
The other parameters are listed in Table \ref{tab:table 7}}.
\label{fig:f9} 
\end {figure}

Fig. \ref{fig:f9} shows effects of various magnetic field strengths on these two lines for 
the persistent phase of 4U 1323--62. We predict that increasing the magnetic field by a small amount changes both 
the line strength and line width. 
Comparing our simulated spectra and the observed data, we estimate the upper limit on the strength of 
prevailing magnetic field in the accretion disc of LMXB 4U 1323--62 to be about 1000 G. 
Here, we determine various physical parameters by only matching the observed flux ratio of Fe XXV and 
Fe XXVI lines at line centre.
%The predicted spectra has not been smoothed.
%We could have better constrain on model parameter had we have  more lines besides Fe XXV and Fe XXVI lines like previous two sources.

\begin{table}
	\centering
	\caption{Predicted physical parameters by CLOUDY modelling of 4U 1323--62.}
	\label{tab:table 7}
	\begin{tabular}{lcr} 
		\hline
		Physical parameters & Persistent & Dip\\
		\hline
%		\begin{deluxetable}{ccc}
%\tablenum{1}
%\tablecaption{Predicted physical parameters by CLOUDY modelling of 4U 1323--62~\label{tab:tab1}.}
%\tablewidth{0pt}
%\tablehead{
%\colhead{Physical parameters} & \colhead{Persistent} & \colhead{Dip}
%}
%\decimalcolnumbers
%\startdata

Power law: slope, log($\xi$) & -1.99, 3.8 & -1.99, 3.3\\
Blackbody: Temperature, log($\xi$) & 10$^{7}$ K, 3.1 & 10$^{7}$ K, 2.5 \\
Density n(r$_{0}$) (cm$^{-3}$) & 10$^{9.3}$ & 10$^{9.3}$ \\
Metallicity & 0.5 solar & 0.5 solar \\
Clump filling & 0.5 & 0.5 \\
Wind (km s$^{-1}$) & 700 & 700 \\
Ionised column density (cm$^{-2}$) & 22.8 & 23.3 \\
Upper limit of disc magnetic field & $\le$ few 100G & -- \\
%\enddata
%\end{deluxetable} 
\hline
\end{tabular}
\end{table}

\subsection{XB 1916--053} \label{subsec:XB 1916--053}
Here we discuss on the last source, XB 1916--053. The observed data for this LMXB are taken from \citet{Iaria2006}. 
%In contrast to the earlier system, 4U 1916--053 does not show Fe XXVI in dipping phase \citep{Juett2005}. 
\citet{Iaria2006} have observed Ne
X, Fe XXV, Fe XXVI, Mg XII, Si XIV and S XVI for XB 1916--053 using $\it Chandra$ satellite. We follow the similar model used for 
GX 13+1. We consider the observed line fluxes reported by
\citet{Iaria2006} and use CLOUDY to model these lines to understand
the underlying physical conditions. In this case, we normalise line fluxes with respect to the weakest Mg XII line flux. 
The wind velocity is not
detected unambiguously here. So, we keep wind velocity as a free
parameter for the model of XB 1916--053. Tabel \ref{tab:table 8} and Table \ref{tab:table 9} list the
best model parameters and predicted line flux ratios using CLOUDY.
 We ran models with three power law radiation spectral indices, -.5, -1.5 and -2. For our
best model, the power law radiation has a spectral index of $-1.5$
and the temperature of the blackbody radiation is $2.8\times
10^{7}$K. The ionisation parameters for power-law and black
body radiations are 3.0 and 1.6, respectively. One can observe that our predicted ionisation parameters and ionised column densities in the persistent 
phase are very close to that 
predicted by \citet{Trigo2006}. However, \citep{Iaria2006} have found a higher value for the ionization parameter, 4.15. 
Our predicted hydrogen density is 10$^{11.77}$ cm$^{-3}$.
Here, the best model predicts that the above mentioned lines originate near the edge of the disc, close to $7.4\times
10^{10}$cm,  and the radial expansion
of the emitting region of these lines to be $10^{11}$ cm. \citet{Iaria2006} have also argued that these lines are produced nearer to $4\times
10^{10}$cm. The predicted
ionised column density is 10$^{22.53}$ cm$^{-2}$. Elemental abundances of
Fe, Mg, S, Si are sub-solar except Ne. All the predicted line flux
ratios are within the observed
range. However,
like the earlier cases, in the final stage
we include a tangled magnetic field in our calculation. We find that with the strength of the magnetic field  greater than $2\times
10^{3}$G, all
the Fe line ratios exceed the observed ranges. No such drastic effect is
observed for Ne X, Si XIV, S XVI lines. We plot our model predictions as a
function of magnetic field for lines Fe XXV and Fe XXVI in Fig. [10]. Comparing our simulated spectra and the observed data, we estimate the upper limit 
on the strength of 
prevailing magnetic field in the accretion disc of XB 1916--053 to be close to 1000 G.

%\begin{deluxetable}{ccc}
%\tablenum{2}
%\tablecaption{Predicted physical parameters by CLOUDY modelling of 4U 1916--053~\label{tab:tab2}.}
%\tablewidth{0pt}
%\tablehead{
%\colhead{Physical parameters} & \colhead{Persistent} & \colhead{Dip}
%}

\begin{table}
	\centering
	\caption{Predicted physical parameters by CLOUDY modelling of XB 1916--053.}
	\label{tab:table 8}
	\begin{tabular}{lr} 
		\hline
		Physical parameters & Best values \\
		\hline
%\decimalcolnumbers
%\startdata
Power law: slope, log($\xi$) & -1.5, 3.0 \\
Blackbody: Temperature, log($\xi$) & $2.8\times 10^{7}$ K, 1.6  \\
Density n(r$_{0}$) (cm$^{-3}$) & $10^{11.77}$  \\
Clump filling & 0.7 \\
Wind (km s$^{-1}$) & 500  \\
Fe/H & -4.170 \\
Mg/H & -5.037 \\
Ne/H & -4.038\\
S/H & -5.250 \\
Si/H & -4.856 \\
Ionised column density (cm$^{-2}$) & 22.53  \\
Upper limit of the disc magnetic field & $\approx$ 1000G  \\
%\enddata
%\end{deluxetable}
\hline
\end{tabular}
\end{table}

\begin{table}
	\centering
	\caption{Comparison of observed and predicted best line flux ratios  of XB 1916--053 using CLOUDY. Lines are normalised w.r.t Mg XII line. }
	\label{tab:table 9}
	\begin{tabular}{lcr} 
		\hline
		Lines $E_{lab}$(kev)  & Observed & Predicted \\		 
		\hline
		Si XIV (2.005)& 0.98--2.39 & 1.59\\
		S XVI (2.621)& 0.64--2.05 & 1.03\\
		Fe XXV (6.700)& 0.49--2.11 & 1.18\\
		Fe XXVI (8.250)& 1.45--4.30 & 1.83\\
		Ne X (1.021) &1.79--5.1 &2.76\\
		\hline
				
	\end{tabular}
\end{table}

\begin{figure}
	% To include a figure from a file named example.*
	% Allowable file formats are eps or ps if compiling using latex
	% or pdf, png, jpg if compiling using pdflatex	
%\plotone{fig1.pdf}

  \vspace{-0.2in}
  \hspace*{0.2in}
  \includegraphics[width=\columnwidth]{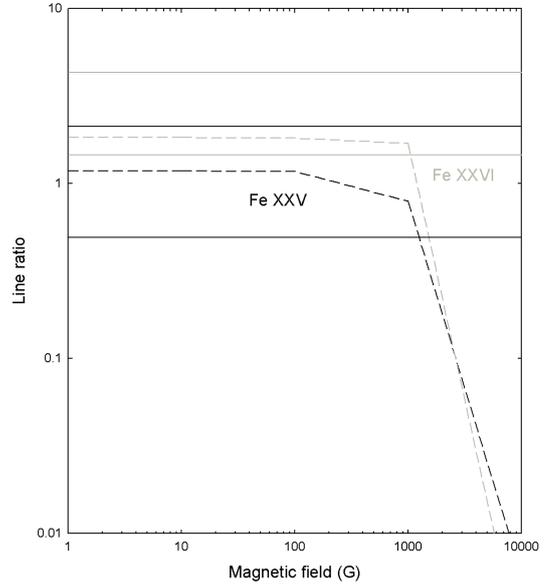}
\vspace*{-.3in}
    \caption{Predicted Fe line ratios for XB 1916--053 is plotted as a function of magnetic field. The horizontal lines represent the observed range (see section 3.4).
    The black and gray lines represent data for Fe XXV and Fe XXVI, respectively.}
    \label{fig:f10}
\end{figure}

\section{Conclusions}
We use the spectral synthesis code CLOUDY \citep{Ferland2013, Ferland2017} to do a detailed modelling of the observed X-ray lines of 
the low-mass X-ray binaries, GX 13+1, MXB 1659--298, 4U 1323--62, XB 1916--053 and determine various underlying physical parameters of these LMXBs. 
In addition to that, we developed a method to estimate an upper limit on the magnetic field of the associated accretion discs by studying these lines.
%including the strength of upper limit of accretion-disc magnetic field.
We assume that the accretion disc has a geometry of a constant-pressure, clumpy, highly-ionised gaseous disc around a 1.4 solar mass neutron star 
with some velocity wind. A rotating disc around a 1.4 solar mass neutron star model prediction matches better with the observed values. In all the models, 
incident radiation is composed of  
both thermal blackbody and a single power law radiation. Our best model calculations predict almost all the line flux ratios within the observed range. Our main conclusions from this work are listed below:
\begin{itemize}
\item All the LMXBs considered here consist of highly ionised accretion discs. Our best models predict the radial extend of the accretion discs from their 
illuminated faces to be close to 10$^{11}$ cm. 
\item Total hydrogen density, consisting of ionised and atomic hydrogen together with all the hydrogen bearing molecules, of the gaseous accretion disc for these four LMXBs spans in the range of $10^{9.3}$ cm$^{-3}$ to $10^{12.1}$ cm$^{-3}$. 
\item We assume that the incident radiation consists of a blackbody radiation and a Comptonised continuum. The temperature of blackbody radiation ranges 
between $1\times 10^{7}$K to $5\times 10^{7}$K. Spectral indices are found to be close to $-2$ for the three LMXBs, MXB 1659-298, 
4U 1323--62.  
XB 1916--053 has a spectral index of -1.5. Whereas, spectral index is close to $-0.6$ for GX 13+1. Similarly, the ionisation parameter for 
the persistent phase of these LMXBs ranges 
between 2.94 to 3.8 for power law radiation and 1 to 3.1 for the blackbody radiation.
\item Ionisation parameters are higher for persistent phases than that of dipping phases.
\item In our models we use the observed wind velocity, whenever available. Otherwise we treat it as a free parameter. The wind velocity for the four LMXBs 
come out to be between 400 km s$^{-1}$ to 700 km s$^{-1}$.
\item For these four LMXBs, most of the heating is contributed by Compton heating.
\item We find that, most of the cooling is contributed by free-free transitions in these four LMXBs.
\item Elemental abundances of Ca and Si are close to their respective solar values but the elemental abundance of Fe is more than its  
solar value for GX 13+1. 4U 1323--62 and XB 1916--053 have sub-solar metallicity (except Ne).
\item We predict presence of clumps in the accretion discs.
\item The predicted ionised column density in the persistent phase for the four LMXBs are between $10^{22.53}$ cm$^{-2}$ to $10^{23.2}$ cm$^{-2}$.
\item Ionised column density is lower in the persistent phases than that of in the dipping phases.
\item We predict an upper limit for the strength of prevailing magnetic field in the accretion discs to be $\sim$ 1000 G which could be verified by future X-ray polarimetric observations.

\end{itemize}

\section{Acknowledgements}

Gargi Shaw acknowledges WOS-A grant from Department of Science and Technology (SR/WOS-A/PM-9/2017). We thank to the referee for valuable comments and 
suggestions.

%%%%%%%%%%%%%%%%%%%%%%%%%%%%%%%%%%%%%%%%%%%%%%%%%%

%%%%%%%%%%%%%%%%%%%% REFERENCES %%%%%%%%%%%%%%%%%%

% The best way to enter references is to use BibTeX:

%\bibliographystyle{mnras}
%\bibliography{example} % if your bibtex file is called example.bib

% Alternatively you could enter them by hand, like this:
% This method is tedious and prone to error if you have lots of references

%%%%%%%%%%%%%%%%%%%%%%%%%%%%%%%%%%%%%%%%%%%%%%%%%%

%%%%%%%%%%%%%%%%% APPENDICES %%%%%%%%%%%%%%%%%%%%%

%\appendix

%\section{Some extra material}

%%%%%%%%%%%%%%%%%%%%%%%%%%%%%%%%%%%%%%%%%%%%%%%%%%

% Don't change these lines
\bsp	% typesetting comment
\label{lastpage}
\end{document}